\documentclass[draft,jgrga]{agutex2015}

\usepackage{url}
\usepackage{mathptmx}
\usepackage{anyfontsize}
\usepackage{t1enc}
\usepackage{amsmath, amssymb, txfonts}
\usepackage[usenames]{color}
\usepackage{threeparttable}
\usepackage{lineno}
\usepackage{lipsum}% http://ctan.org/pkg/lipsum
\usepackage{multicol}% http://ctan.org/pkg/multicols
  \usepackage[dvipdf]{graphicx}

  \setkeys{Gin}{draft=false}
\authorrunninghead{ESMAEILI ET AL.}

\titlerunninghead{Wavefunction Properties of a Single}

\begin{document}

%% ------------------------------------------------------------------------ %%
%
%  TITLE
%
%% ------------------------------------------------------------------------ %%

\title{%[Estimation of]
Wavefunction Properties of a Single and a System of Magnetic Flux Tube(s) Oscillations
%Or
%[Estimation of] probability distribution function of extreme magnetic storms
}
%
% e.g., \title{Terrestrial ring current:
% Origin, formation, and decay $\alpha\beta\Gamma\Delta$}
%

%% ------------------------------------------------------------------------ %%
%
%  AUTHORS AND AFFILIATIONS
%
%% ------------------------------------------------------------------------ %%

%Use \author{\altaffilmark{}} and \altaffiltext{}

% \altaffilmark will produce footnote;
% matching \altaffiltext will appear at bottom of page.

% \authors{R. C. Bales,\altaffilmark{1}
% E. Mosley-Thompson,\altaffilmark{2} R. Williams,\altaffilmark{3}
% J. R. McConnell\altaffilmark{4}, and Francesco Visconti\altaffilmark{5}}

\authors{S. Esmaeili \altaffilmark{1}
M. Nasiri, \altaffilmark{1}
N. Dadashi , \altaffilmark{1}
H. Safari \altaffilmark{1}
}

\altaffiltext{1}{Physics Department, University Blvd., University of Zanjan, Zanjan, I. R. Iran,  P. Box. 45371-38791.
(shahriar.esmaeili@znu.ac.ir)}

%\altaffiltext{1}{Department of Hydrology and Water Resources,
%University of Arizona, Tucson, Arizona, USA.}

%\altaffiltext{2}{Department of Geography, Ohio State University,
%Columbus, Ohio, USA.}

%\altaffiltext{3}{Department of Space Sciences, University of
%Michigan, Ann Arbor, Michigan, USA.}

%\altaffiltext{4}{Division of Hydrologic Sciences, Desert Research
%Institute, Reno, Nevada, USA.}

%\altaffiltext{5}{Dipartimento di Idraulica, Trasporti ed
%Infrastrutture Civili, Politecnico di Torino, Turin, Italy.}
\keypoints{Finite element method is applied for solving the magnetoacoustic wave
equation for a single and a system of flux tubes.
Properties of wavefunctions for various modes denoted by
azimuthal, longitudinal, and radial numbers are
investigated for a flux tube.
Properties of wavefunctions are studied in a system of flux tubes cases. }
%% ------------------------------------------------------------------------ %%
%
%  ABSTRACT
%
%% ------------------------------------------------------------------------ %%

% >> Do NOT include any \begin...\end commands within
% >> the body of the abstract.

\begin{abstract}

In this study, the properties of wavefunctions of the MHD oscillations for a single and a system of straight flux tubes are investigated. Magnetic flux tubes with a straight magnetic field and longitudinal density stratification were considered in zero-$\beta$ approximation. A single three-dimensional wave equation (eigen value problem) is solved for longitudinal component of the perturbed magnetic field using the finite element method (FEM). Wavefunctions (eigenfunction of wave equation) of the MHD oscillations are categorized into sausage, kink, helical kink and fluting modes. Concerning the amplitude location of the waves that arise from the flux tube, those waves are identified as body, surface, and leaky waves, appeared in the case of both a single and a system of flux tubes. Exact recognition of the wavefunctions can be used in coronal seismology and also helps to the future high resolution instruments that would be designed for studying the properties of the solar loop oscillations in details.

\end{abstract}

%% ------------------------------------------------------------------------ %%
%
%  BEGIN ARTICLE
%
%% ------------------------------------------------------------------------ %%

% The body of the article must start with a \begin{article} command
%
% \end{article} must follow the references section, before the figures
%  and tables.

%% ------------------------------------------------------------------------ %%
%
%  TEXT
%
%% ------------------------------------------------------------------------ %%

\begin{article}
\section{Introduction}

 Wave propagation in the magnetic flux tubes is one of the most fundamental problems of MHD theory structured as a building block to infer the physics behind the several solar phenomena \citep{Priest1}. Since launching the first instruments for observing the sun, many different wave modes have been detected in the solar corona (standing and propagating slow- and fast-modes) which is discussed in a new field named coronal seismology. To obtain the physical properties of the coronal loops that are not easy to measure directly, the properties of the waves (such as the magnetic field, heating rate, and fine structure parameters) could be used. The original idea of wave properties in a nonuniform medium was suggested by \cite{Uchida} and extended by \cite{Roberts1} for slab and cylindrical geometries. An extended review on  the theoretical and observational properties of the coronal flux tube oscillations is given by \cite{Nakariakov2}.

 In the past decades, the special high-resolution observations taken by TRACE, SoHO, Yohkoh, etc. were applied to detect the kink oscillations in coronal loops \citep{Aschwanden1, Nakariakov1, Aschwanden2, Schrijver1, wang+solanki}. Theoretical investigations such as the tube curvature, damping time, mode profiles, and the oscillations frequencies  were carried out to depict a real physical image of the coronal loops \citep{Gruszecki1, Ruderman1, van Doorsselaere1, Verwichte1, Dymova+Ruderman}. The oscillatory behavior of the longitudinally stratified cylindrical loops are studied by \cite{Andries1, Safari1, Erdelyi+verth}. The effects of the internal structure on the properties of transverse oscillations were investigated by \cite{Arregui1}. \cite{Pascoe1} studied the effects of a nonuniform cross section on the standing sausage modes of a coronal loop. Magnetic tubes with the elliptic cross-sections were analysed by \cite{Ruderman2}. \cite{verth+Erdelyi} assessed the effect of both the density and magnetic stratification on the transverse oscillations of a loop. The nature of kink MHD waves in magnetic flux tube was discussed by \cite{Goossens2}.

 It seems that in the helically twisted loops, torsional oscillation modes should be measurable. The global torsional oscillations and Alfv\'{e}n waves with long wave length may be detectable through the spectral line width along the loop and the spatial variation of the Doppler shift, respectively. The variation of global torsional modes in Fe~{\small\small{\bf XIV}} coronal emission line spectra in the upper layers of an active region at the solar limb were detected by \cite{zaqarashvili1}. \cite{Portier-Fozzani} suggested that the spatial motion along the loop with suitable helical geometries inferred for spatial verification of torsional MHD modes \citep{Aschwanden3}.

Comparing the frequencies of the collective oscillations of loops covering similar kink frequencies with those of the individual kink modes were studied by \cite{Luna1}. Transverse oscillations of a multi-stranded coronal loops described as several parallel cylindrical strands are investigated by \cite{Luna2}. They concluded the transverse oscillations of a loop get influence by the internal fine structure and the equivalent monolithic loop cannot be properly described by its transverse dynamics. The method of \cite{Luna1} was generalized by \cite{Fathalian+Safari} on the collective kink-like modes of the system of coronal loops to involve longitudinal density stratification along the loop axis.

The current research aims to model the identifications based on spatial behavior of wavefunction of straight single and a system of flux tubes. The governed linearized MHD equations are reduced to a single partial differential equation by applying the boundary conditions for the total pressure and the normal velocity component \citep{Edwin+roberts, Sakurai1, Goossens1}. Both eigenfrequencies and wavefunctions are extracted to identify the modes of MHD oscillations.

The paper is organized as follows: In Section \ref{GEB}, we introduce the problem as a wave equation and discuss the boundary conditions governing on. In Section \ref{Methods}, we focus on the wavefunction to obtain numerical solutions and graphical results for both single and a system of flux tubes. Further, we discuss about our numerical method and graphical results which are given by extracting the wavefunctions. In Section \ref{Res} we review the categorization of wave propagation in the nonuniform medium with our numerical solutions. The model is generalized for a system of flux tubes that are located randomly inside its equivalent tube. The wavefunctions of flux tubes are studied in two cases of identical and nonidentical system of flux tubes. In Section \ref{Sum}, summary and discussion are given.

\section{Governing Equations and Boundary Conditions}\label{GEB}
\subsection{Governing Equations}\label{GE}

\begin{figure}
\centerline{\includegraphics[width=1.08\textwidth,clip=]{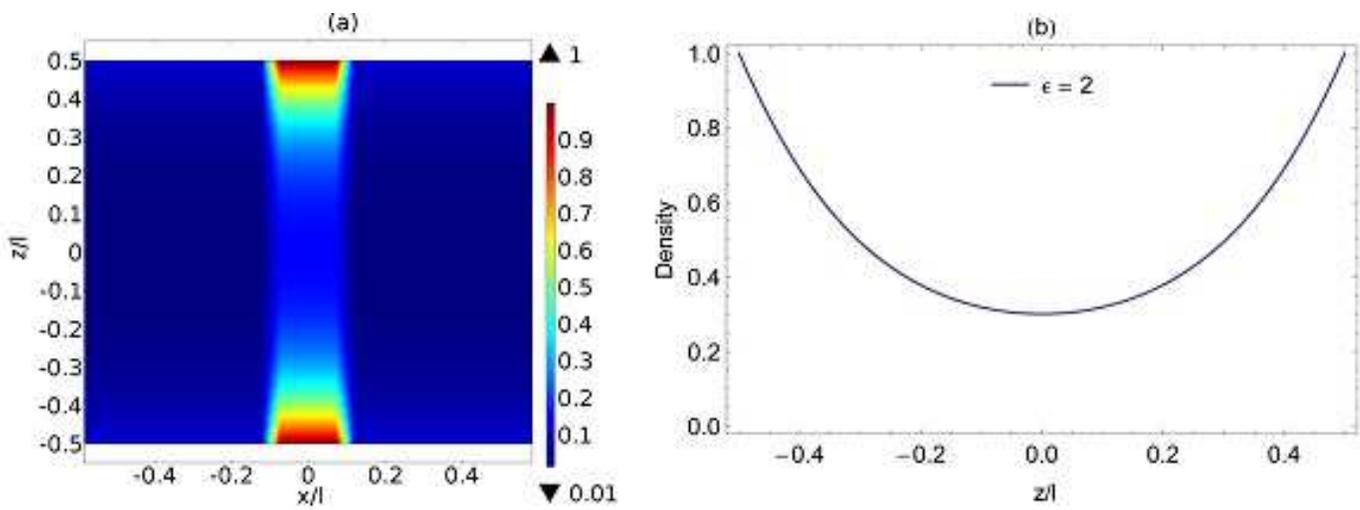}}
\caption{a: Contour plot of the density profile shown in the $x-z$ plane with the stratification parameter $\epsilon = 2$ along the loop axis. b: The density profile along the $z$ axis at $x = 0$.}\label{fig1}       % Give a unique label
\end{figure}

MHD wave propagation in a spatially structured straight flux tube pervaded by magnetically straight plasma medium is studied. The equilibrium magnetic field is assumed to be straight and constant along the flux tube axis, $\mathbf{B} = B\hat{z}$. The density for the inside and outside the tube is $\rho_{i}$ and  $\rho_{e}$, respectively. To make a longitudinal structuring, the density profile is assumed to be stratified by the function $f(\epsilon, z) = \exp(-\epsilon \cos\frac{\pi z}{l})$, the stratification parameter is \textbf{$\epsilon= l/\pi H$}, where $H$ indicates the atmospheric scale height.  Thus, the equilibrium density profile for a single flux tube with its environment taking the form of

\begin{equation}\label{eq1}
\rho(\vec{r}, \epsilon) = \left((\rho_{i} - \rho_{e}) \Theta \left(a-\sqrt{x^2+y^2}\right) + \rho_{e}\right) f(\epsilon, z),
\end{equation}
where the density variation between inside and outside the flux tube is demonstrated by a step function $(\Theta)$ across the lateral surface. The length and radius of the flux tube are $l$ and $a$, respectively. The tube ends are frozen at \textbf{$z = \pm l/2$} due to a dense photospheric plasma so that the density has a minimum at $z= 0$ and two maxima at \textbf{$z = \pm l/2$} (Fig. \ref{fig1}). The gravity forces and all dissipative terms are neglected.

For a system of \textbf{$N$} flux tubes, the density profile could be

\begin{equation}\label{eq2}
\rho(\vec{r}, \epsilon) = \sum_{j=1}^N\left((\rho_{i,j} - \rho_{e}) \Theta\left(a-\sqrt{x^2_{j}+y^2_{j}}\right) + \rho_{e}\right) f(\epsilon, z),
\end{equation}
where the subscript $j$ indicates the flux tube number.

To describe the linear motion, we consider a small amplitude perturbation proportional to $exp(-i\omega t)$ for all perturbed quantities. Therefore, the governing linearized MHD equations are reduced to the following single partial differential equation that denotes a wave equation in zero-$\beta$ approximation:
\begin{equation}\label{eq3}
\nabla^2 b_{z}(r, \theta, z) + \frac{\omega^2}{\text{v}^2_{A}(r, \theta, z, \epsilon)}b_{z}(r, \theta, z) = 0.
\end{equation}
where $\text{v}^2_{A} = B^2/4\pi\rho(\epsilon, z)$ is the square of the local Alfv\'{e}n speed and $b_{z}$ is the perturbed magnetic field in the $z$ component \citep{Fathalian+Safari}. In order to use zero-$\beta$ condition, the slow modes are removed from our studies.

\subsection{Boundary Conditions}\label{BC}
Equation (\ref{eq3}) is  an eigenvalue problem and can be solved in cylindrical coordinates with appropriate boundary conditions. The electrical field can be written in the form of  $\mathbf{E} = -{\bf v} \times {\bf B}$ in ideal MHD equations. Regarding that the tube ends are bounded in the photosphere, ${\bf E} = 0$, the tangential component of the vector  ${\bf v} \times {\bf B}$ is zero at the tube ends \citep{Ruderman+Goossens}
\begin{equation}\label{eq4}
{\bf e}_{z} \times ({\bf v} \times {\bf B}) = 0, \qquad {\rm at} \qquad z = \pm l/2.
\end{equation}

In order to consider the standing modes of the tube oscillations in $z$ direction, the perturbed quantities (v and $P_{T}$) are considered as
\begin{equation}\label{eq5}
\text{v}_{\perp}= 0 \qquad P_{T} =0 \qquad {\rm at} \qquad z =\pm l/2,
\end{equation}
where $P_{T}$ is defined as the magnetic pressures. In our model, both the interior and exterior magnetic fields are assumed to be constant but distinct. Thus, the boundary conditions for total pressure and the normal velocity component are kept in balance for inside and outside of the flux tube. Otherwise,  the waves may steepen into shock at the lateral surface.
	
The boundary conditions at the lateral surface of flux tube $(r = a)$ are give by \citep[e.g.][]{Edwin+roberts, Donnelly1}
\begin{equation}\label{eq6}
{\bf n} \cdot \lbrack {\bf v}\rbrack = 0, \qquad {\bf n} \cdot  \lbrack{\bf B}\rbrack = 0, \qquad \lbrack P_{T} \rbrack = \lbrack \frac{\mathbf{B} \cdot \mathbf{b_{z}}}{8\pi}\rbrack =0.
\end{equation}
The Bracket represents the difference across the tube boundaries. At the tube axis, $r = 0$, we assume
\begin{equation}\label{eq7}
    b_{z}(r=0, \theta, z) = {\rm finite},
\end{equation}
and at large distance from the tube axis we supposed evanescent condition
\begin{equation}\label{eq8}
b_{z}(r\to \infty, \theta, z)\to 0.
\end{equation}
Equation (\ref{eq3}) can be solved numerically for interior and exterior of the flux tube concerning with an appropriate density profile along the tube and its medium by imposing the boundary conditions (Eqs. \ref{eq5}-\ref{eq8}).

\section{Solution Using the Finite Element Method}\label{Methods}
\begin{figure}
\centerline{\includegraphics[width=0.9\textwidth,clip=]{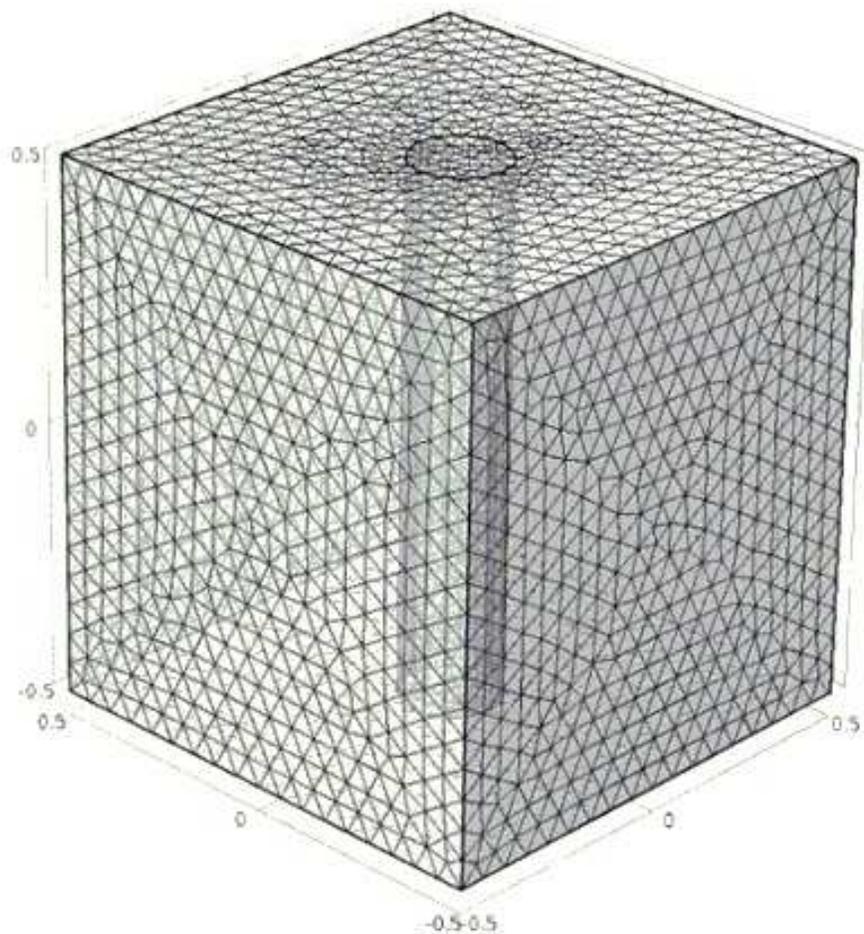}}
\caption{Meshed cube by finite element method with tetrahedral domain representing a flux tube bounded at $z = \pm 0.5$ inside it.}\label{fig2}       % Give a unique label
\end{figure}

We consider a geometry composed of longitudinally stratified flux tube inside a bounded cube in the cylindrical coordinates to solve numerically Eq. (\ref{eq3}). The employed method to solve our partial differential equation, Eq. (\ref{eq3}), is the finite element method (FEM).  The primitive variables, eigenfrequencies, and the  wavefunctions are computed. In this method, each domain is divided into a collection of sub-domains. Each sub-domain represented by a set of element equations to the original problem. The domains type to perform the (FEM) solution is free tetrahedral. Figure \ref{fig2} represents  a sketch of a cylinder in a meshed bounded cube made by FEM.  For more details about the FEM see \cite{strang+fix}.

\section{Results}\label{Res}
Using  finite element method, Eq. (\ref{eq3}) is numerically solved by imposing the boundary conditions and the described model. In our numerical solution for a single flux tube, the ratio between the radius and cylinder length is chosen to be $a/l = 0.1$. The density ratio for the inside and outside is equal to $\rho_{e}/\rho_{i} = 0.1$. The oscillations frequencies are dimensionalized to fundamental angular Alfv\'{e}n frequency $(\omega_{A}=2~rad/s)$.

As stated in literatures \citep[e.g.][]{Edwin+roberts, Goossens1}, the oscillations of a single uniform and untwisted flux tube are categorized into sausage $m=0$, kink $m=1$ and fluting modes $m = 2$. Hearafter, the mode numbers $m$, $k$, and $n$ are used for the number of antinodes along the azimuthal component $(\theta)$, the tube axis, and the radial direction, respectively \citep{Karami}.

 Sausage modes are described in a geometry including symmetric treatment on the tube cross section. In other words, a local decrease in the radial direction due to a slight perturbation increases the strength of the local magnetic field, and then it is suppressed by the inward magnetic pressure which leads to sausage mode. It can be inferred in Fig. \ref{fig3} that the tube cross sections involve in a symmetric fluctuation and the lateral surface of the tube oscillates in opposite directions. In Fig. \ref{fig3} (a), the wavefunction of fundamental sausage mode $(m = 0,~k = 1,~n=1)$ is presented. As it is shown in Fig. \ref{fig3} (b), by increasing the stratification parameter ($\epsilon$) the antinode is getting flattened along the tube axis and moves toward the tube ends. Fig. \ref{fig3} (c) represents the normalized line profile along the solid line plotted in Fig. \ref{fig3} (a) and (b) at $x = 0$ for various stratification parameters.

Kink modes describes the transverse displacement of the cross section owing to the slight perturbation in a same direction \citep{Goossens2, Ofman, Ruderman+Goossens}.  The first overtone kink mode $(m = 1,~k = 2,~n=1)$ is represented in Fig. \ref{fig4}. Since the antinodes asymmetrically appear around the tube axis, it is interpreted as kink oscillations. The antinodes are similarly move toward the tube ends by increasing the stratification parameters $(\epsilon)$, Fig. \ref{fig4} (b).  The line graph of the wavefunction for different stratification parameter are plotted in Fig. \ref{fig4} (c).

Torsional modes give rise to the oscillations at the internal Alfv\'{e}n speed $(\text{v}_{Ai})$. These modes are azimuthally symmetric, so that the cross section and the tube boundaries are not displaced transversally, hence it cannot be directly detected by observation \citep{Nakariakov2, van Doorsselaere2}. The term related to torsional Alfv\'{e}n waves were removed in our equations. We here extract the modes which relate to the magnetoacoustic waves. Therefore, the wavefunctions of a mode are interpreted as helical kink mode (Figs. \ref{fig5}, and \ref{fig6}). This mode is a type of kink modes which also possess twisted behaviors inside the flux tube. The wavefunction of $x-z$ plane is in accord with the fundamental kink mode shown in Fig. \ref{fig5}. The wavefunction of this mode is further plotted for different cross sections along the tube axis in Fig. \ref{fig6} . The line slopes between the locations of the antinodes (indicated by strikes) are obtained for the various cross sections from $z = -0.4$ to $z = 0.4$. As it is shown, the slope is altered by the height $(z)$.  Movie 1 shows the twist of antinodes inside the flux tube that moves from the lower to the upper cross sections along the tube axis.

 The wavefunction of fluting mode $(m=2,~k=2,~n=1)$ including four antinodes inside the tube is also represented by our numerical solution, (Fig. \ref{fig7}, a). Corresponding wavefunction in $x-z$ plane is plotted in Fig. \ref{fig7} (b).

%The wavefunctions and oscillation modes are computed for different stratification parameters.  \textbf{Fundamental sausage mode $(m = 0,~k = 1)$ and first overtone kink mode $(m = 1,~k = 2)$}  are illustrated. Normalized line graph of two dimensional wavefunctions are plotted along a line consists of two anti-nodes. By changing the stratification parameter, the antinodes move toward the tube ends \textbf{ \citep{Erdelyi+verth, Safari1, Karami}}, see Figs. \ref{fig3} and \ref{fig4}.

Following \cite{Edwin+roberts}, the analytical solution of the derived dispersion relation can be written by Bessel functions. Body, Surface, and leaky waves are described by different Bessel functions that are the solution of MHD wave equation. We also extracted these waves in two and one dimensional cases.  Figure  \ref{fig8} represents the wavefunction of a body mode. As shown in Fig. \ref{fig8} (b), the maximum amplitudes is located inside the tube.
 Wavefunction of the fundamental body mode (surface) is demonstrated in Fig. \ref{fig9}.  The maximum amplitudes of this type of waves occur at the lateral surface (interface), the line graph along the inclined line is also plotted in Fig. \ref{fig9} (b). Leaky waves are a type of waves arise in their exterior medium, the maximum amplitude takes place outside the tube. In Fig. \ref{fig10}, the wavefunction of leaky mode is represented. It is clear that the maximum amplitude forms outside the tube, (Fig \ref{fig10}, b).

In the next step, we generalize our numerical solution for a system of $9$ flux tubes that are distributed randomly inside a hypothetical flux tube. The wavefunctions of the oscillations can be used to interpret the modes. Each thin tube with length $l$, radius $a = 0.01 l$, and the internal density $\rho_{i,j}$($j$ indicates the tube number) is located in a flux tube with radius $R$. The thin tubes are wrapped with a hypothetical flux tube of radius $(R = 0.065 l)$ in a way that their center positions $(\mathbf{r} = x_{j}\hat{x} + y_{j}\hat{y})$ generate randomly inside the  hypothetical flux tube. Condition $d \geq (2a) + 0.01$ is applied for the center positions of the circles. The term $0.01$ denotes the minimum distance among the center of the circles. The density of the tube environment is $\rho_{e}$ (Fig. \ref{fig11}). Following \cite{Luna2}, for a system of $N$ thin tubes, Eq. (\ref{eq9}) is obtained for the density of equivalent monolithic tube,
\begin{equation}\label{eq9}
\rho_{eq} =  \sum_{j=1}^N \rho_{i,j}(\frac{a}{R})^2 + \rho_{e}\left(1 - N\left(\frac{a}{R}\right)^2\right).
\end{equation}
Equation (\ref{eq3}) is numerically solved for a system of $N = 9$ identical tubes, concerning the density profile is taken Eq. \ref{eq2}. Each tube has identical density and radius. The density of each tube is assumed to be $\rho_{i,j} = 7.5~\rho_{e}$, with radius $a$, thus the equivalent density is obtained $\rho_{eq} = 9.519 \rho_{e}$ from Eq. \ref{eq9}. Wavefunctions of kink-like modes for a system of 9 identical thin tubes and the fundamental kink mode for the equivalent monolithic loop are investigated. Figure \ref{fig12} shows the wavefunctions of the both a system of identical tubes and equivalent monolithic loop in $x-y$ and $x-z$ planes.
Moreover, in the case of nonidentical system, the density ratio of the tubes are taken to be $\rho_{i}/\rho_{e}=\{7.89,~7.61,~7.60,~8.97,~5.98,~8.73,~7.52,~8.62,~6.18\}$, and for the equivalent tube yields $\rho_{eq}~=~9.580~\rho_{e}$ from Eq. \ref{eq9}. Likewise, the wavefunctions of the system of nonidentical tubes and its equivalent monolithic tube by applying the stratification parameter are extracted, (Fig. \ref{fig13}).

On the basis of our numerical solution, the wavefunctions of the collective kink-like oscillations in both identical and nonidentical systems are compared with those of kink oscillations related to their equivalent monolithic tubes. The results show that both the systems of identical and nonidentical flux tubes have approximately the same wavefunctions with the equivalent monolithic tube.

\section{Summary and Discussion}\label{Sum}
We have studied the MHD oscillations of a single and a system of magnetized flux tube(s) under the zero-$\beta$ conditions. A single equation (Eq. \ref{eq3}), for $z$ component of the perturbed magnetic field, is numerically solved based on finite element method.

\begin{figure}
\centerline{\includegraphics[width=1.08\textwidth,clip=]{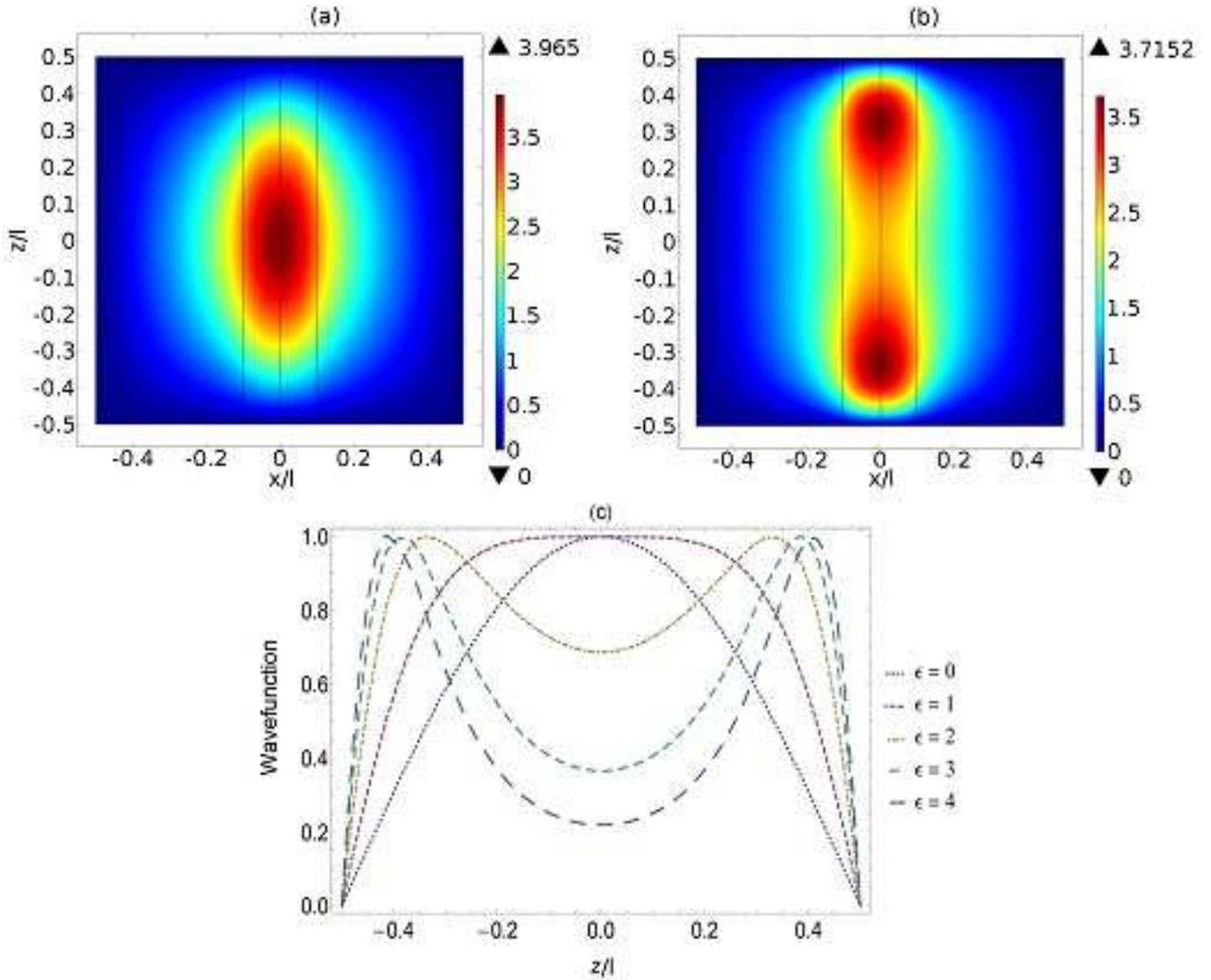}}
\caption{a: Two dimensional wavefunction in the $x-z$ plane at the point of $y = 0$ representing the fundamental sausage mode $(m = 0,~k = 1, n=1)$ with the stratification parameter $\epsilon = 0$. b: The corresponding wavefunction with $\epsilon = 2$. c: Normalized line profile of two dimensional wavefunctions including maximum antinode along the solid line at $x = 0$ represented in panels $a$ and $b$ for various stratification parameters.}\label{fig3}       % Give a unique label
\end{figure}

\begin{figure}
\centerline{\includegraphics[width=1.08\textwidth,clip=]{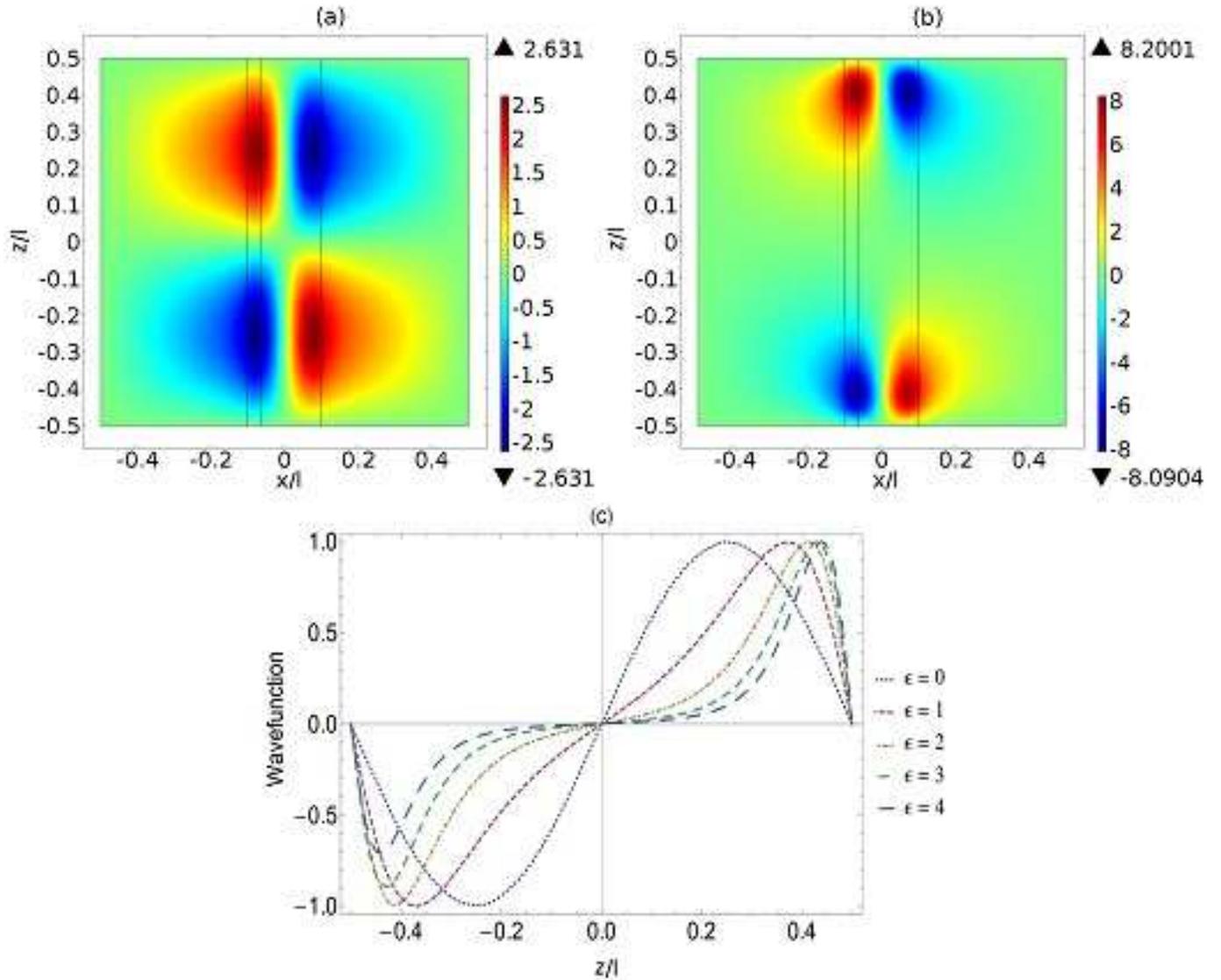}}
\caption{a: Two dimensional wavefunction in the $x-z$ plane at the point of $y = 0$ representing the first overtone kink mode $(m = 1,~k = 2, n=1)$ with the stratification parameter $\epsilon = 0$. b: The corresponding wavefunction with $\epsilon = 2$. c: Normalized line graph of two dimensional wavefunctions including maximum antinode along the solid line at $x = -0.680$ represented in panel $a$ and $b$ for various stratification parameters.} \label{fig4}
\end{figure}

\begin{figure}
\centerline{\includegraphics[width=.8\textwidth,clip=]{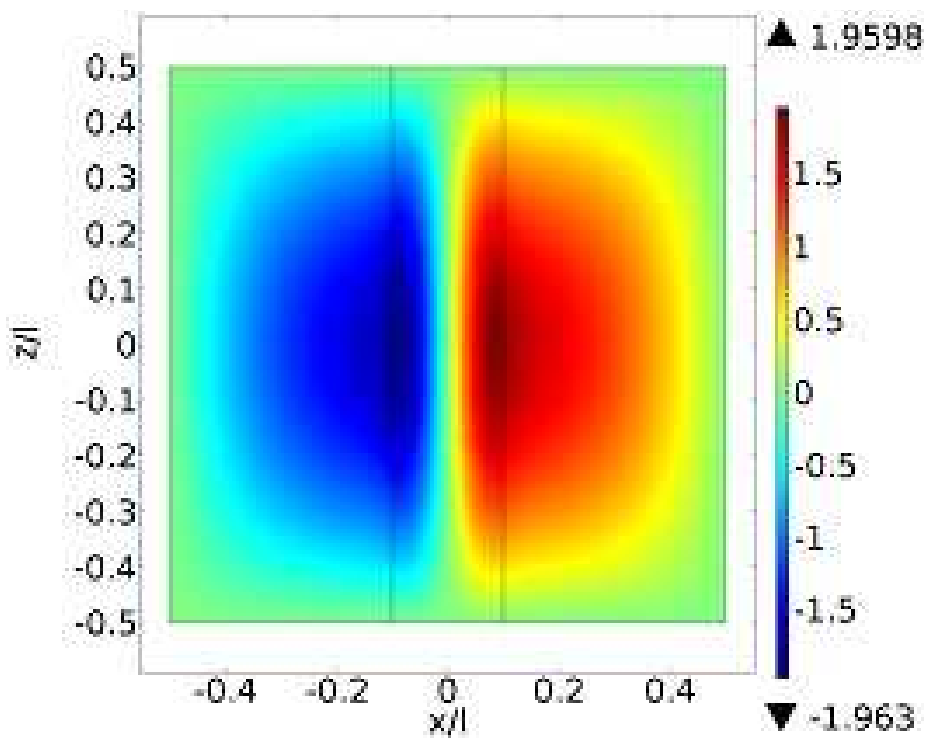}}
\caption{Two dimensional wavefunction which is interpreted as helical kink mode represented in $x-z$ plane. Its helical behavior is indicated in Fig. \ref{fig6}.}\label{fig5}       % Give a unique label
\end{figure}

\begin{figure}
\centerline{\includegraphics[width=1.1\textwidth,clip=]{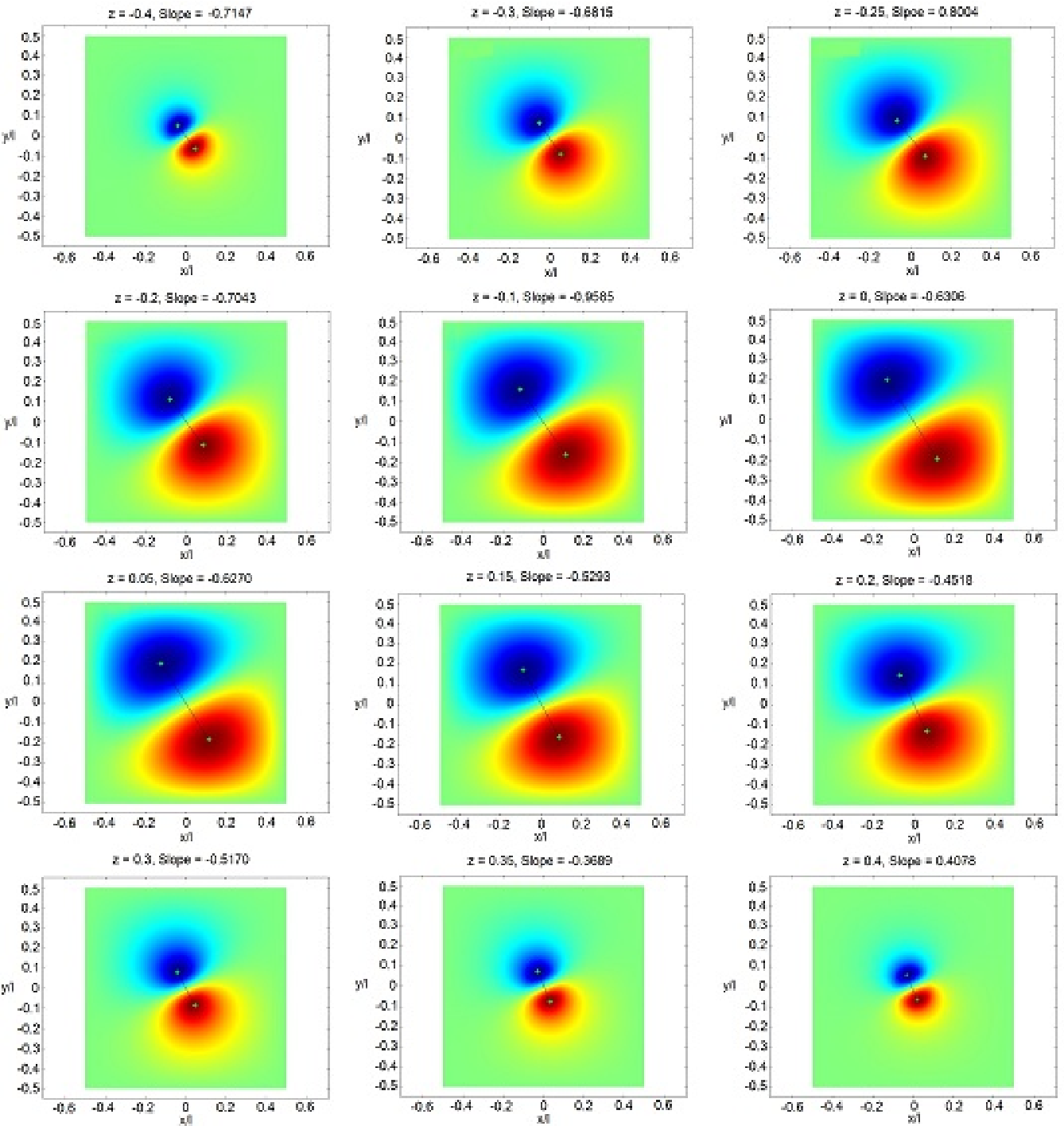}}
\caption{Representing the  wavefunctions of Eq. (\ref{eq3}) for different cross sections along the tube axis. The line slope between the maximum and minimum points vary along the tube axis. It can be concluded as helical behavior inside the flux tube and introduced as helical kink mode.}\label{fig6}       % Give a unique label
\end{figure}

\begin{figure}
\centerline{\includegraphics[width=1.07\textwidth,clip=]{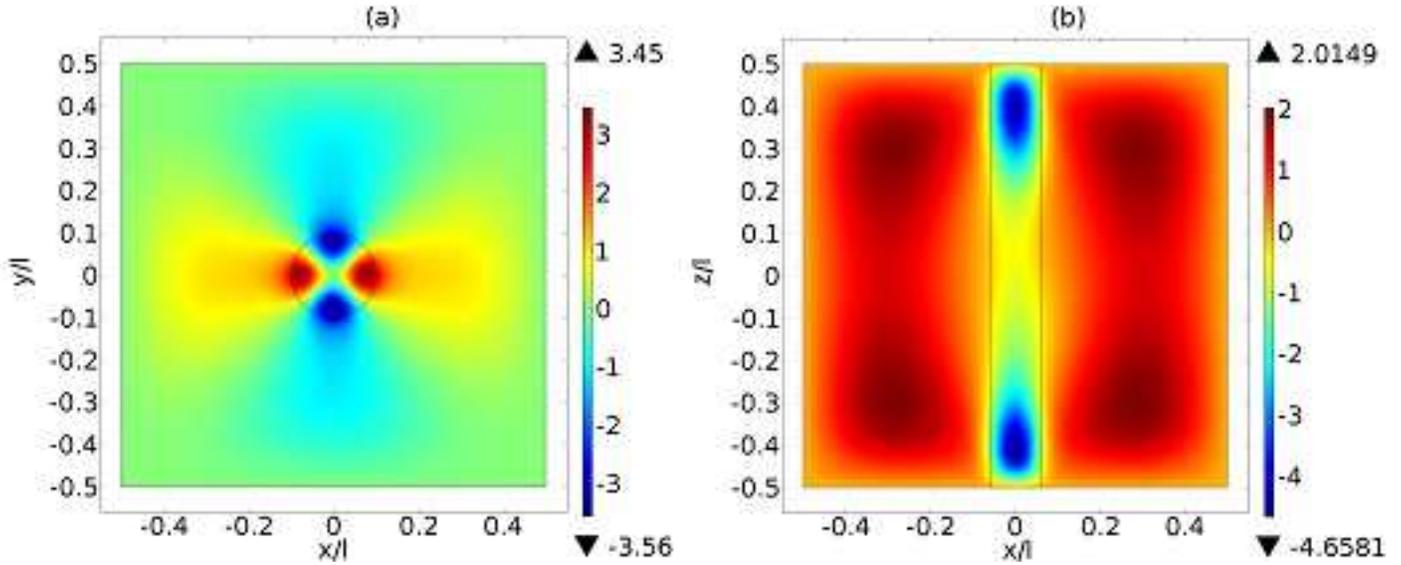}}
\caption{a: Two dimensional wavefunction in the $x-y$ plane at the point of $z = 0$ including four antinodes inside the tube representing the first overtone fluting mode $(m = 2,~k = 2, n=1)$ with the stratification parameter $\epsilon = 1$. b: The corresponding wavefunction in the $x-z$ plane at the point of $y = 0$ with $\epsilon = 1$.} \label{fig7}
\end{figure}

The properties of the wavefunctions of sausage, kink, helical kink, and fluting modes are extracted. We see the antinodes are moving toward the tube ends by changing the stratification parameters $(\epsilon)$ (Figs. \ref{fig3} and \ref{fig4}).  We detect a type of kink mode in which twisted along the tube axis and we interpret it  as a helical kink mode (Fig. \ref{fig5} and \ref{fig6}). Sice the line slope between the location of antinodes is changed along the tube axis, that shows a twisting inside flux tube (Movie. 1). To future this goal, we are exploring the detection of torsional modes in our numerical solution domain. If successful, the torsion mode analysis would also be applied to observational data for composition of future instruments designed for studying the solar loops.

\begin{figure}
\centerline{\includegraphics[width=1.12\textwidth,clip=]{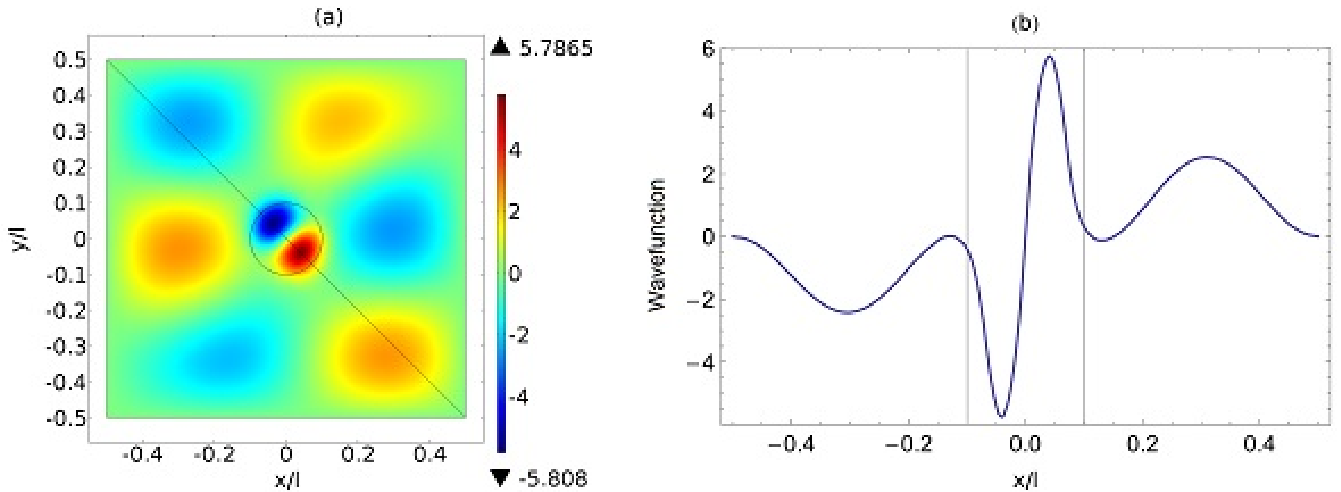}}
\caption{a: Two-dimensional wavefunction of Eq. (\ref{eq3}) in the $x-y$ plane represents the body wave propagation in a nonuniform plasma medium. b: Representng the wavefunctions versus $x/l$ along the solid inclined line shown in the left panel. Maximum amplitude exists inside the tube boundaries shown by the $2$ solid lines.}\label{fig8}       % Give a unique label
\end{figure}

\begin{figure}
\centerline{\includegraphics[width=1.12\textwidth,clip=]{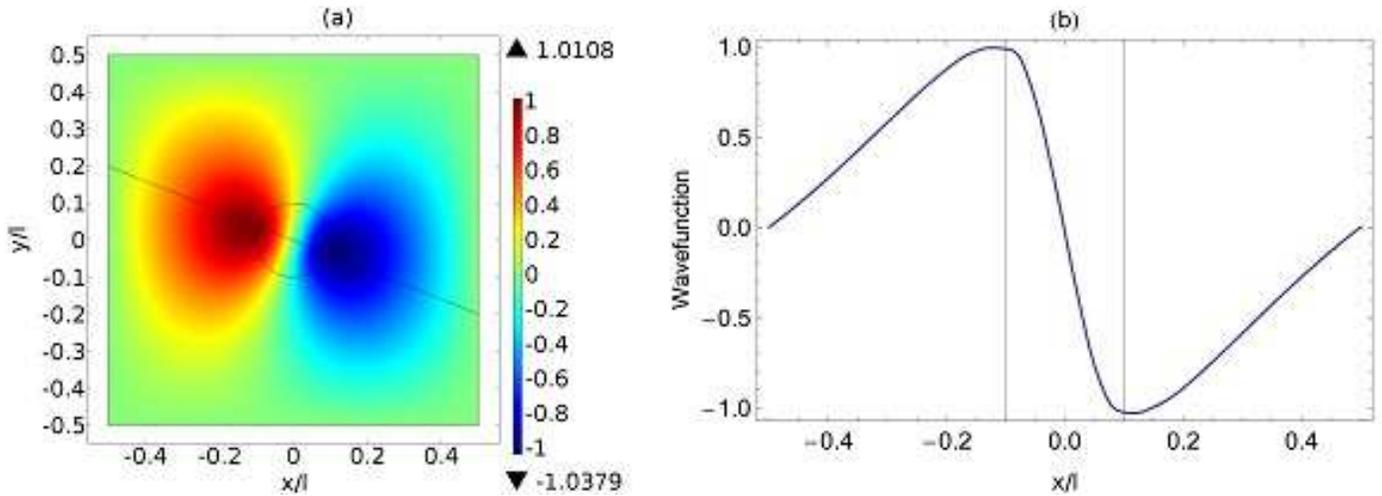}}
\caption{a: Two-dimensional wavefunction of Eq. (\ref{eq3}) in the $x-y$ plane represents the fundamental body (surface) wave propagation in a nonuniform plasma medium. b: Representing the wavefunctions versus $x/l$ along the solid inclined line shown in the left panel. Maximum amplitude located at the lateral surface of flux tube shown by $2$ solid lines.}\label{fig9}       % Give a unique label
\end{figure}

\begin{figure}
\centerline{\includegraphics[width=1.12\textwidth,clip=]{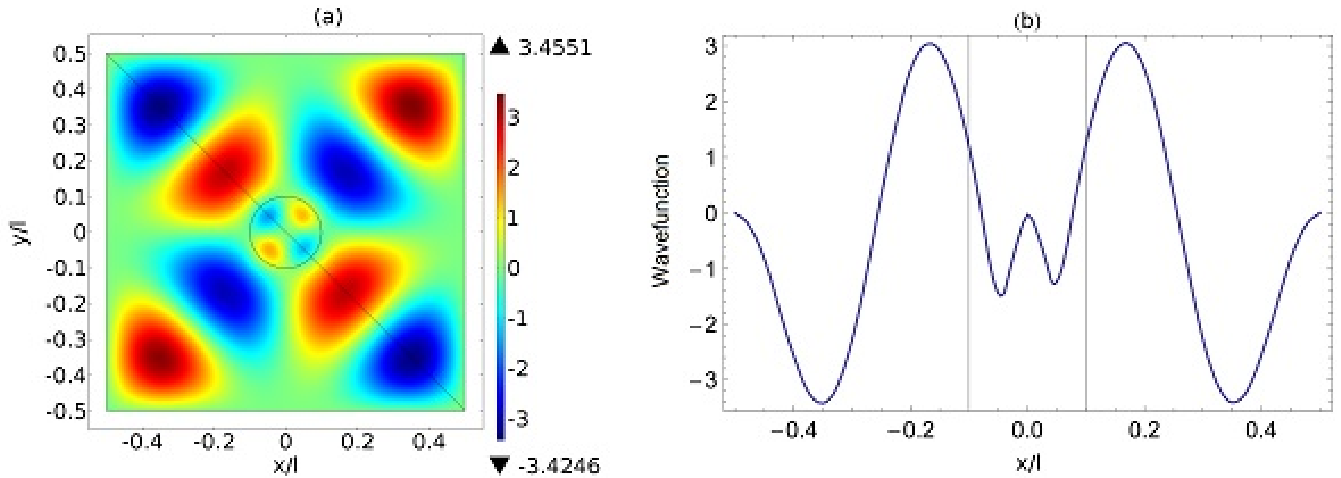}}
\caption{a: Two-dimensional wavefunction of Eq. (\ref{eq3}) in the $x-y$ plane represents the leaky wave propagation in a nonuniform plasma medium. b: Representing the wavefunctions versus $x/l$ along the solid inclined line shown in the left panel. Maximum amplitude exists outside the tube boundaries shown by $2$ solid lines.}\label{fig10}       % Give a unique label
\end{figure}

\begin{figure}

\centerline{\includegraphics[width=1.1\textwidth,clip=]{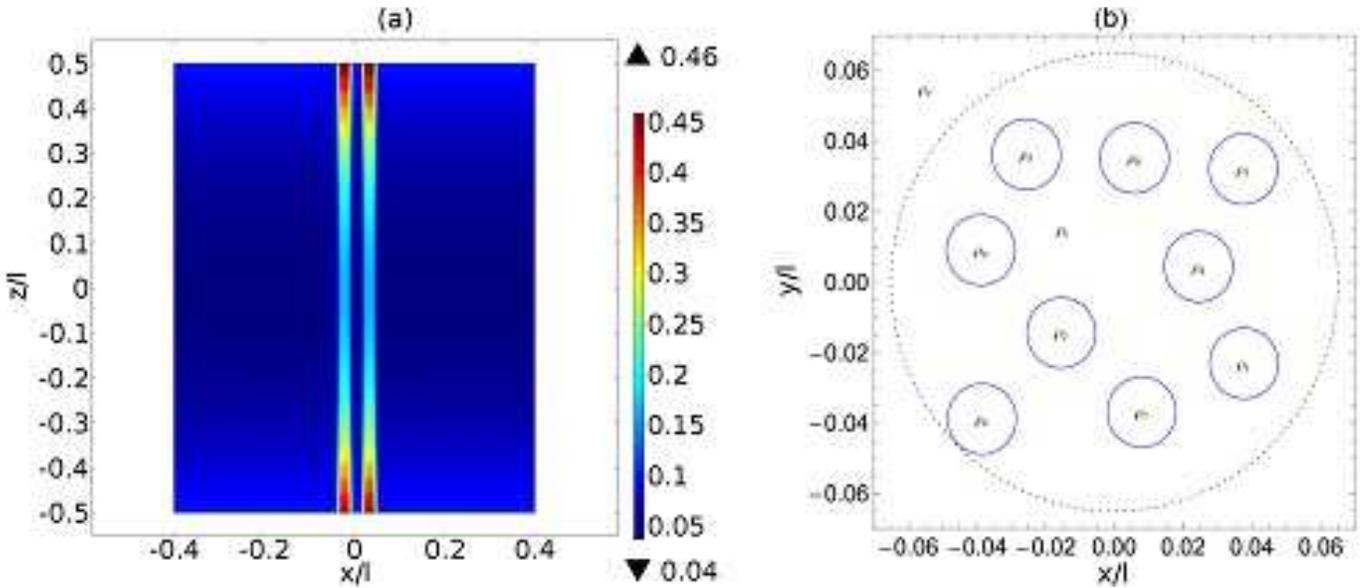}}
\caption{a: Contour plot of the density profile for a system of flux tubes shown in the $x-z$ plane at $y = -0.038$ with the stratification parameter $\epsilon = 1$ along the loop axis. b: A sketch of the cross section of a system of $9$ thin flux tubes where the circles are distributed randomly under the condition of ($d \geq 2a +  0.01$). The density and radius of thin tubes are  $\rho_{i}$ and $a = 0.01 l$ (blue smaller circles), respectively, which is inscribed by a circle of radius $R$ (dotted circle) displaying the equivalent tube. The density of external medium and the medium between the tubes is $\rho_{e}$.}\label{fig11}       % Give a unique label
\end{figure}

\begin{figure}
\centerline{\includegraphics[width=1\textwidth,clip=]{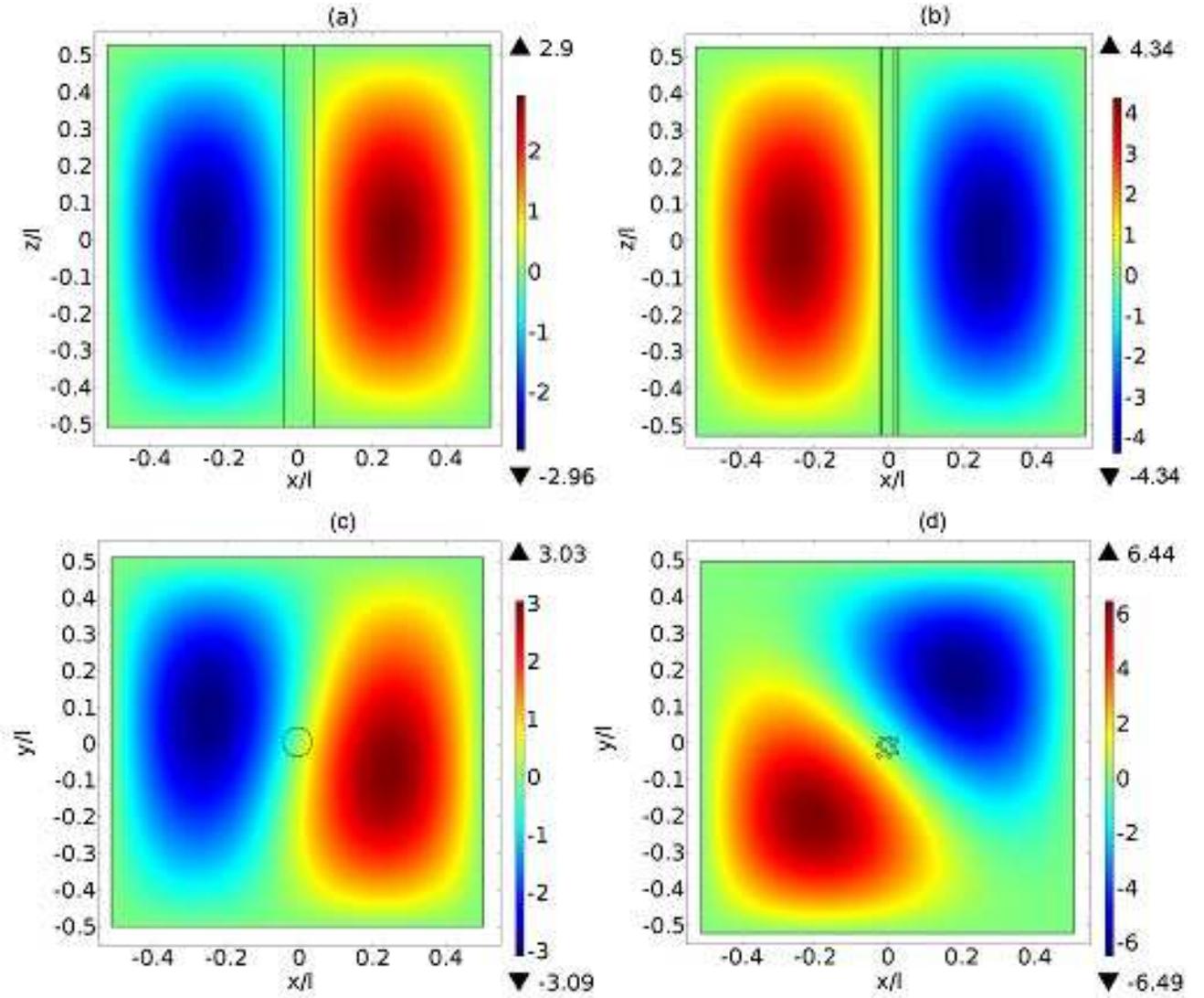}}
\caption{a: Representing fundamental kink mode of the hypothetical monolithic flux tube with radius $R = 0.065 l$ related to the system of $9$ identical thin tubes shown in the $x-z$ plane at $y = 0$ with the stratification parameter $\epsilon = 0$. The boundaries of the monolithic flux tube are denoted by two solid lines. b: Two dimensional wavefunction in $x-z$ plane at $y=0$ for a system of $9$ thin flux tubes. The boundaries of $2$ thin tubes located at $y = 0$, (shown in Fig. \ref{fig11}, b), denoted by four solid lines. The wavefunction is related to kink-like oscillations. Panels (c) and (d) demonstrate the corresponding wavefunction in the $x-y$ plane at $z = 0$ for the equivalent monolithic tube and a system of $9$ identical thin tubes, respectively.} \label{fig12}       % Give a unique label
\end{figure}

\begin{figure}
\centerline{\includegraphics[width=1\textwidth,clip=]{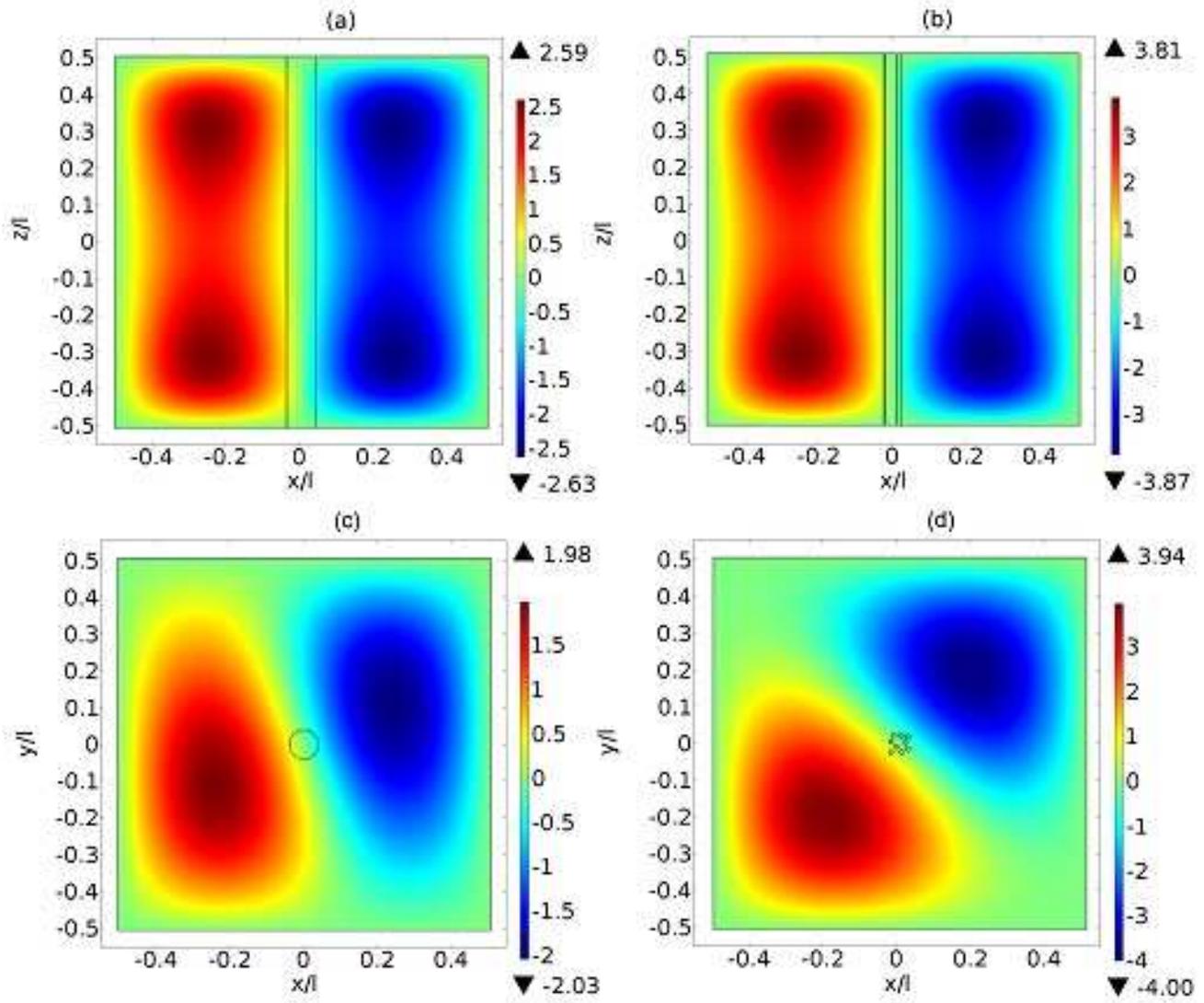}}
\caption{a: Two dimensional wavefunction of the hypothetical monolithic flux tube with radius $R = 0.065 l$ related to the system of $9$ nonidentical thin tubes shown in the $x-z$ plane at $y = 0$ with the stratification parameter $\epsilon = 2$ along the tube axis. The boundaries of the monolithic flux tube are denoted by the $2$ solid lines. b: Two dimensional wavefunction in the $x-z$ plane at $y=0$ for a system of $9$ thin flux tubes. The boundaries of $2$ thin tubes located at $y = 0$, (shown in Fig. \ref{fig11}, b), denoted by $4$ solid lines. Panels (c) and (d) demonstrate the corresponding wavefunction  in the $x-y$ plane at $z = 0$ for the equivalent monolithic tube and a system of $9$ nonidentical thin tubes, respectively. Panels (a) and (c) show the fundamental kink mode, for the oscillations of system of tubes it is called kink-like oscillations, panels (b) and (d).} \label{fig13}       % Give a unique label
\end{figure}

In other classification of wave propagation, those are categorised into body, surface, and leaky with respect to their positions of maximum amplitude in the nonuniform medium. Moreover, we extracted the wavefunctions of such waves from our numerical solution shown in Figs.  \ref{fig8}, \ref{fig9} and  \ref{fig10}. The numerical solution for both a system of identical and nonidentical flux tubes was developed. The collective kink-like oscillations of a system of both identical and nonidentical tubes were studied. In the stratified system of flux tubes, we concluded from our wavefunctions that the frequencies of kink-like modes are approximately the same as the kink modes of an individual monolithic flux tube, (see Figs. \ref{fig12} and  \ref{fig13}) \citep{Luna1, Luna2, Fathalian+Safari}.

\begin{acknowledgments}
S. Esmaeili is grateful to Prof. Jim Klimchuk and NSF for their support for participating at the Triennial Earth-Sun Summit 2015 conference. Our numerical code is available from the authors upon request (shahriar.esmaeili@znu.ac.ir).
\end{acknowledgments}

%Reference citation examples:

%...as shown by \textit{Kilby} [2008].
%...as shown by {\textit  {Lewin}} [1976], {\textit  {Carson}} [1986], {\textit  {Bartholdy and Billi}} [2002], and {\textit  {Rinaldi}} [2003].
%...has been shown [\textit{Kilby et al.}, 2008].
%...has been shown [{\textit  {Lewin}}, 1976; {\textit  {Carson}}, 1986; {\textit  {Bartholdy and Billi}}, 2002; {\textit  {Rinaldi}}, 2003].

%...as shown by \citet{jskilby}.
%...as shown by \citet{lewin76}, \citet{carson86}, \citet{bartoldy02}, and \citet{rinaldi03}.
%...has been shown \citep{jskilbye}.
%...has been shown \citep{lewin76,carson86,bartoldy02,rinaldi03}.
%
% Please use ONLY \citet and \citep for reference citations.
% DO NOT use other cite commands (e.g., \cite, \citeyear, \nocite, \citealp, etc.).

%% ------------------------------------------------------------------------ %%
%
%  END ARTICLE
%
%% ------------------------------------------------------------------------ %%

\end{article}

%% Enter Figures and Tables here:

% When submitting articles through the GEMS system:
% COMMENT OUT ANY COMMANDS THAT INCLUDE GRAPHICS.
%
% FOR FIGURES, DO NOT USE \psfrag or \subfigure commands.
%
% Figure captions go below the figure.
% Table titles go above tables; all other caption information
%  should be placed in footnotes below the table.

% DRAFT figure/table, including eps graphics
%
% \begin{figure}
% \noindent\includegraphics[width=20pc]{samplefigure.eps}
% \caption{Caption text here}
% \end{figure}
% \end{document}
%
% \begin{table}
% \caption{}
% \end{table}
%
% ---------------
% TWO-COLUMN figure/table
%
% \begin{figure*}
% \noindent\includegraphics[width=39pc]{samplefigure.eps}
% \caption{Caption text here}
% \end{figure*}
%
% \begin{table*}
% \caption{Caption text here}
% \end{table*}

% Figure 1
\newpage
% Figure 2


%\bibliographystyle{spr-mp-nameyear-cnd}
%%\bibliography{myref}
%\bibliography{References}
%\nocite{*}

\end{document}